\documentclass[runningheads]{llncs}
\usepackage{graphicx}
\usepackage{comment}
\usepackage{tcolorbox}
\usepackage{url}
\usepackage{xcolor}
\usepackage{verbatim}
\usepackage{caption}
\usepackage{multirow}
\usepackage{booktabs} % For better-looking tables
\usepackage{makecell} % For line breaks in table cells
\usepackage{siunitx}  % For alignment (optional here as your data is text)
\usepackage{caption}
\usepackage{subcaption}
\usepackage{amsmath} % For math symbols
\usepackage{tabularx}
\usepackage{geometry}
\usepackage{todonotes} % TODO

\newcommand{\modelname}{\textit{S3LLM}}
\newcommand{\llmname}{LLaMA-2}

\begin{document}

\title{{\modelname}: Large-\textit{S}cale \textit{S}cientific \textit{S}oftware Understanding with \textit{LLMs} using Source, Metadata, and Document}

\titlerunning{\modelname}

\author{
     Kareem Shaik\inst{1} \and 
     Dali Wang\inst{2} \and 
     Weijian Zheng\inst{3} \and 
     Qinglei Cao\inst{4} \and \\
     Heng Fan\inst{1} \and 
     Peter Schwartz\inst{2} \and 
     Yunhe Feng\inst{1}
% \orcidID{0000-0001-6577-227X}
}

\authorrunning{K. Shaik et al.}
\institute{University of North Texas, Denton, TX 76207, USA \\ \email{kareembabashaik@my.unt.edu,\{heng.fan, yunhe.feng\}@unt.edu}\\ \and
Oak Ridge National Laboratory, Oak Ridge, TN 37830, USA \email{\{wangd,schwartzpd\}@ornl.gov} \and 
Argonne National Laboratory, Lemont, IL 60439, USA \email{wzheng@anl.gov} \and Saint Louis University, St. Louis, MO 63103, USA \email{qinglei.cao@slu.edu}}

\maketitle  % typeset the header of the contribution

\begin{abstract}

The understanding of large-scale scientific software poses significant challenges due to its diverse codebase, extensive code length, and target computing architectures. The emergence of generative AI, specifically large language models (LLMs), provides novel pathways for understanding such complex scientific codes. This paper presents {\modelname}, an LLM-based framework designed to enable the examination of source code, code metadata, and summarized information in conjunction with textual technical reports in an interactive, conversational manner through a user-friendly interface. {\modelname} leverages open-source {\llmname} models to enhance code analysis through the automatic transformation of natural language queries into domain-specific language (DSL) queries. Specifically, it translates these queries into Feature Query Language (FQL), enabling efficient scanning and parsing of entire code repositories. In addition, \textit{\modelname} is equipped to handle diverse metadata types, including DOT, SQL, and customized formats. Furthermore, {\modelname} incorporates retrieval augmented generation (RAG) and LangChain technologies to directly query extensive documents. {\modelname} demonstrates the potential of using locally deployed open-source LLMs for the rapid understanding of large-scale scientific computing software, eliminating the need for extensive coding expertise, and thereby making the process more efficient and effective.
S3LLM is available at \textcolor{blue}{\url{https://github.com/ResponsibleAILab/s3llm}}.

\end{abstract}


\section{Introduction}

Large-scale scientific computing software is crucial in various scientific fields, undergoing extensive development cycles that lead to the formation of intricate software libraries and ecosystems. This complexity stems from the lengthy development periods, ongoing extensions, and evolving development paradigms, making it imperative to provide users with insights into these computing tools. However, understanding such software is a challenging task due to several factors. First, large-scale scientific software often incorporates multiple programming languages, including older languages such as Fortran and Pascal, which poses a significant challenge for contemporary programmers trying to understand the code. Second, the large volume of scientific software, which may encompass millions of lines of code, presents the obstacle to comprehensively understanding each segment of the code. Lastly, the documentation for these software systems is sometimes less than ideal, often lacking detailed explanations, which further complicates the task of gaining a thorough understanding of the software.

To enhance comprehension of large-scale scientific software, numerous tools have been devised to aid in code analysis and documentation. For instance, Doxygen \cite{doxygen} is capable of generating documentation and performing static code analysis for software source trees. Similarly, Sphinx \cite{sphinx} is compatible with a wide range of programming languages, making it especially effective for producing exhaustive documentation across various formats, including HTML, LaTeX (for printable PDF versions), ePub, Texinfo, manual pages, and plain text. Nonetheless, the currently available tools are primarily tailored for static code analysis and lack the capability to accommodate dynamic queries. Moreover, given the complexity inherent in large-scale scientific software, it poses a significant challenge for both developers and users to formulate queries in both instructed (e.g., textural documents) and structured formats (e.g., SQL). Thus, it is imperative to devise methods for understanding and parsing large-scale scientific software that are both user-friendly and precise.

The emergence of generative AI, particularly large language models (LLMs), heralds a new era in software comprehension and interaction. LLMs have shown remarkable capabilities across various tasks, including chatbot interactions \cite{llm-chat1,llm-chat2,llm-chat3}, text summarization \cite{llm-sum1,llm-sum2,llm-sum3}, and content creation \cite{llm-gen1,llm-gen2,llm-gen3}, demonstrating their potential to revolutionize programming and documentation practices. Beyond these applications, LLMs offer promising solutions for navigating and understanding the complex landscapes of large-scale scientific software \cite{llm-scientific1}. By leveraging LLMs, we can envision a future where software comprehension is not only more accessible but also more intuitive, enabling users to query and interact with software in natural language. This paper introduces {\modelname}, a novel framework that embodies this vision, providing a user-friendly interface for interacting with complex scientific computing software through conversational, natural language queries. {\modelname} aims to bridge the gap between the intricate world of scientific software and the diverse community of users and developers, fostering a deeper understanding and facilitating more effective use of these critical computational tools.

Different from most existing works on software understanding, the proposed {\modelname} can handle various types of tasks for large-scale scientific software understanding including source code query, metadata analysis, and text-based technical report understanding. {\modelname} is capable of conducting queries over the information extracted from source code in diverse formats, such as DOT (graph description language)\footnote{\url{https://en.wikipedia.org/wiki/DOT_(graph_description_language)}} and relational database. By leveraging the few-shot learning capability of LLMs, {\modelname} can also generate domain-specific language (DSL) queries, such as Feature Query Language (FQL) ~\cite{FQL}, to gather and extract software features through code analysis. Furthermore, {\modelname} implements LangChain and Retrieval-Augmented Generation (RAG)~\cite{lewis2020retrieval} schemes to enable text-based queries from technical reports and project summaries. More importantly, all aforementioned interactions and inquiries facilitated by {\modelname} are executed utilizing natural language.

The contributions of this paper are summarized as follows: 
\begin{itemize}
\item We have conceptualized, designed, and implemented {\modelname}, a novel framework that utilizes LLMs to improve the understanding of large-scale scientific software. This framework excels in analyzing source code, metadata, and textual technical reports, providing a holistic approach to software comprehension.
\item {\modelname} presents a user-friendly interface that employs natural language processing, allowing users, even those with limited programming knowledge, to easily query and gain insights into scientific software.
\item Recognizing the need to balance inference speed with the framework's computational demands, {\modelname} provides three options featuring {\llmname} models with 7B, 13B, and 70B parameters, allowing users to choose the most appropriate model based on their specific requirements.
\item Experiments conducted with the large-scale Energy Exascale Earth System Model (E3SM)~\cite{e3sm} demonstrate the effectiveness of our model in analyzing source code, metadata, and textual documents.
\item We contribute to the scientific computing community by releasing {\modelname} as an open-source tool, ensuring broad accessibility and usefulness across a broad spectrum of scientific computing applications and research pursuits.
\end{itemize}

\section{Related Work}

\textbf{Code information collection:} A variety of tools have been developed to gather diverse forms of code information. Tools such as cloc~\cite{cloc}, sloc~\cite{sloc}, and sonar~\cite{sonar} are designed to assess a project's source code to determine its size and the programming languages employed. Meanwhile, the ScanCode \cite{scancode} toolkit and fossology \cite{fossology} are specialized tools that provide insights on software licenses, copyrights, dependencies, and additional relevant information. The OSS Review Toolkit~\cite{ossreviewtoolkit} further enhances these capabilities by integrating third-party package managers (e.g., MAVEN, PIP, NPM) and code scanners (e.g., Licensee, ScanCode), facilitating the identification of dependencies across different open-source libraries within a project.
Nonetheless, these tools do not leverage Artificial Intelligence (AI) or LLMs to simplify the process of collecting code information, which requires users to input their requirements in a format that these tools can interpret. In contrast, our framework {\modelname} distinguishes itself by allowing users to express their requirements in natural language, subsequently providing precise and accurate code information. This approach significantly improves the efficiency and accuracy of code analysis.

\noindent\textbf{LLM-based Software Engineering:} The integration of AI and LLMs has significantly transformed code analysis and software development methodologies. This evolution is evident in the widespread application of LLMs for code generation, underscoring their utility in enhancing programming efficiency and accuracy \cite{feng2023investigating,dong2023self}. Furthermore, research in this domain has validated the effectiveness of LLMs in critical tasks such as unit test generation~\cite{unittest1,unittest2}, bug analysis~\cite{bugana1,bugana2}, and debugging~\cite{debug1,debug2}, showcasing their potential to refine testing processes, improve bug detection, and streamline debugging.
Recently, researchers have also applied LLM to large-scale scientific software~\cite{llm-scientific1}. However, instead of employing LLM to understand the scientific code, researchers typically focus on specific tasks such as extracting variables of interest by reading the code documentation~\cite{llm-scientific1}.
Despite the prevalent focus on code generation, testing, and repair, {\modelname} diverges by leveraging both LLM and traditional techniques to deepen the understanding of large-scale scientific codes.

\section{Method}

We first introduce an overview of the {\modelname} framework, subsequently delving into the detailed design of its components. {\modelname} examines large-scale scientific software from multifaceted perspectives, utilizing diverse data types such as source code, code metadata, and textual reports. Each component's design is meticulously outlined to elucidate how {\modelname} facilitates a comprehensive investigation of scientific software, ensuring a thorough understanding of its complex ecosystem.

\subsection{Framework Overview}

\begin{figure}[t]
    \centering
    \includegraphics[width=\columnwidth]{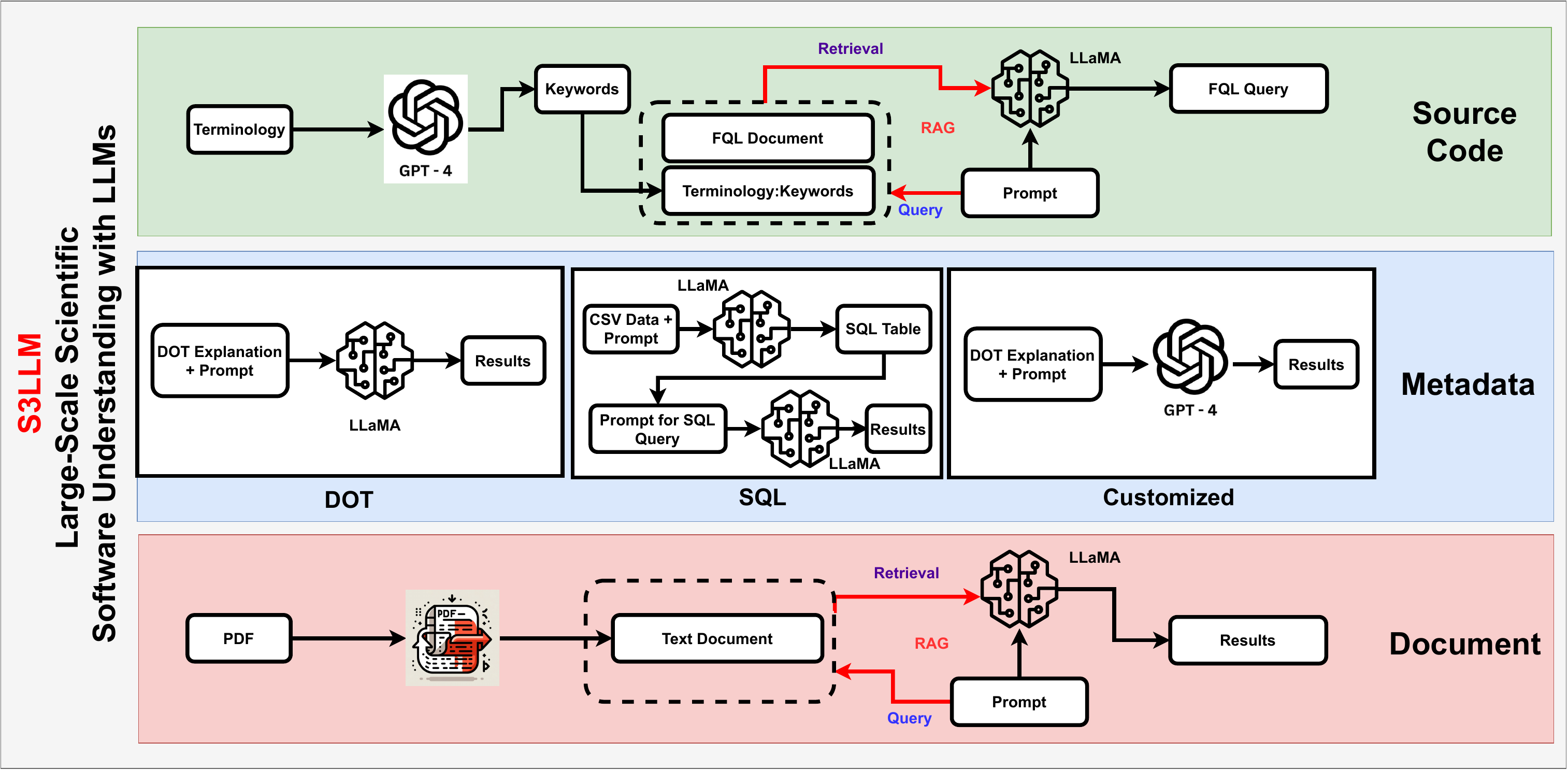}
    \caption{Framework overview of {\modelname} 
    }
    \label{fig:framework-overview}
\end{figure}

At the heart of {\modelname} lies open-source {\llmname} models engineered for conversational interactions in natural language. {\llmname} has been intricately configured to support multi-round conversations, maintaining awareness of the ongoing context to ensure continuity and relevance in its responses, thus embodying in-context learning capabilities. As depicted in Figure~\ref{fig:framework-overview}, {\modelname} is architecturally composed of three primary components dedicated to processing source code, code metadata, and textual technical documents. For all of the three components, {\llmname} models play a key role in translating natural languages into desired domain-specific language (DSL) queries or in text information analysis and retrieval. In addition, some consisting components in {\modelname} adopt RAG, a methodology designed to enhance the responses of an LLM by consulting an external and authoritative knowledge base not included within its initial training data, thereby refining its output prior to generating a response.
The choice of open-source {\llmname} models (7B, 13B, and 70B) as the foundation for all language understanding tasks within {\modelname} is strategic, offering versatility in model sizes and ensuring a high degree of reproducibility across various computational environments.

\subsection{Source Code Analysis}

Large-scale scientific software, such as E3SM, often contains vast quantities of source code, exceeding a million lines in some cases. This presents a significant challenge for open-source LLMs in directly managing such a huge number of tokens by loading all of the tokens into their context windows. Furthermore, many sophisticated source code analysis tools today require users to possess in-depth programming expertise or specialized domain knowledge for effective source code interrogation. To address these issues, {\modelname} combines the code analysis capability of existing tools and the natural language understanding capability of LLMs together, not only avoiding loading the entire codebase into LLMs but also allowing users to query source code using natural language.

In illustrating the strategy of {\modelname} for LLM-based source code analysis, we highlight XScan~\cite{zheng2019xscan} as a representative backend engine. XScan is an integrated software toolkit designed to extract key information of large-scale scientific code, such as lines of code, programming languages, external library dependencies, and architecture-dependent parallel software features. While XScan provides user-friendly methodologies for basic source code analysis tasks (e.g., adopting Doxygen to construct caller-callee graphs or using CLOC~\cite{cloc} to tally lines of code), it also introduces FQL for more nuanced static code analysis. Despite XScan offering straightforward solutions for basic analysis through simple command executions, FQL's reliance on user-generated queries to investigate software features represents a significant hurdle. As Table~\ref{tbl:example_fql_types} demonstrates, FQL accommodates various query types, including Library Utilization, Version Assessment, and Feature Enumeration Queries, all of which demand a degree of coding proficiency from the user for precise query formulation. {\modelname} is specifically designed to bridge this gap, addressing the critical challenge of facilitating accessible source code analysis without requiring extensive programming knowledge. 
It is important to highlight that the core LLMs powering {\modelname} are broadly applicable and capable of integrating with source code analysis engines beyond just XScan.

\begin{table}[t]
	\centering
 	\centering
	\captionsetup{justification=centering}
	\caption{Examples of HPC feature questions and associated FQL queries}
	\label{tbl:example_fql_types}
\begin{tabular}{|c|c|}
\hline
\textbf{Question} & \textbf{FQL Query} \\ \hline
\begin{tabular}[c]{@{}c@{}}\textit{Is OpenMP used?} \\ \textbf{Library Utilization Query}\end{tabular}                                    & \texttt{FQL: CHECK (!\$OMP $\|$ pragma omp) WHERE (*) AS (OpenMP)} \\ \hline
\begin{tabular}[c]{@{}c@{}} \textit{What is the minimum} \\ \textit{version requirement of MPI?} \\ \textbf{Version Assessment Query} \end{tabular} & \begin{tabular}[c]{@{}c@{}} \texttt{MAX (CHECK (MPI\_AINT\_ADD $\|$ MPI\_AINT\_DIFF) WHERE (*) AS (3.1),}\\ \texttt{...,} \\ \texttt{CHECK (mpi.h $\|$ use mpi $\|$ mpif.h) WHERE (*) AS (2.0))}\end{tabular} \\ \hline
\begin{tabular}[c]{@{}c@{}}\textit{What OpenMP scheduling}\\ \textit{method is used?} \\ \textbf{Feature Enumeration Query} \end{tabular} & \begin{tabular}[c]{@{}c@{}} \texttt{LIST (CHECK (schedule(static) WHERE(*) AS (Static),} \\ \texttt{CHECK (schedule(dynamic) WHERE(*) AS (Dynamic), ...,} \\ \texttt{CHECK (schedule(runtime) WHERE(*) AS (Runtime))}\end{tabular} \\ \hline
\end{tabular}
\vspace{-0.5cm}
\end{table}

In the conversion process from natural language to FQL queries, {\modelname} must master three crucial elements: the purpose and syntax of FQL, comprehension of specific terminologies (e.g. high-performance computing (HPC) programming standards) presented in natural language inputs, and the establishment of an accurate translation from natural language to FQL. To equip the LLM models within {\modelname} with a comprehensive grasp of FQL and its programming syntax, we have incorporated a foundational document on FQL alongside a collection of FQL query examples into the RAG framework as external data sources. For a detailed exploration of the RAG methodology utilized in this study, please refer to Section~\ref{sec:doc-rag}. Addressing the conversion of relevant terminologies from natural language inputs into programming code keywords posed a significant challenge. Initial attempts to generate these keywords using open-source LLM models yielded suboptimal results. Consequently, we select GPT-4 to serve as the terminology translator within {\modelname}. This approach simplifies the creation of mappings from terminologies to corresponding programming keywords with precision. These mappings are also stored as external data within RAG. Finally, we prompt {\modelname} to generate new FQL queries from provided natural language questions by utilizing few-shot learning techniques and enriching the context with RAG data.

\subsection{Software Metadata Comprehension}

In the quest to deepen our comprehension of large-scale scientific software, a variety of code metadata is extracted utilizing current software analysis tools. Particularly, metadata related to software architecture and data structures stands out as some of the most valuable and enlightening for understanding software intricacies. To capitalize on this rich metadata, {\modelname} has been meticulously crafted not only to interpret these data but also to respond to queries informed by them. Fundamentally, our framework is adept at managing well-structured metadata formats, such as DOT, thereby enhancing its utility in parsing and understanding the underlying structure and organization of software code. 

To enhance {\modelname}'s capability to handle diverse information from scientific computing software, we implement a dual-phase strategy. Initially, we ensure that the LLMs grasp the structure and organization of metadata formats. This is achieved by integrating context rich in such information, utilizing either the RAG technique or by directly embedding metadata format information into the prompts. Subsequently, {\modelname} allows users to interrogate the metadata through natural language prompts in such a context. Notably, {\modelname} introduces two distinct prompting methodologies: zero-shot and few-shot modes. The zero-shot mode enables users to perform queries on metadata without requiring prior example inputs, whereas the few-shot mode solicits demonstration examples from users to refine and guide the LLM's responses. The few-shot will be more useful when processing very complex software metadata. 

{\modelname} currently supports three types of metadata extracted from large-scale scientific software: DOT, SQL, and specified data formats by third-party software. However, it is designed with the flexibility to easily accommodate additional metadata formats in the future.
For the DOT format, we facilitate its understanding in \textit{\modelname} by incorporating a detailed explanation of DOT within the prompt, followed by appending the specific query question.
Given the widespread use of SQL in relational database management, we leverage the \textit{\llmname} models to generate SQL queries directly, without the need for instruction-based prompting techniques.
For custom data formats specified by third-party software, we use the bespoke metadata produced by SPEL~\cite{schwartz2022spel}, a toolkit developed for adapting E3SM models for GPU execution via OpenACC, as a case study. This choice illustrates {\modelname}'s capability to interpret highly specialized data formats through the application of GPT-4, underscoring our commitment to extending {\modelname}'s utility to encompass a broad spectrum of metadata types, including those tailored by third-party software analysis tools.

\subsection{Technical Document Interpretation}~\label{sec:doc-rag}
Large-scale scientific software often comes with a comprehensive set of supplementary documentation, such as technical reports, user manuals, and research papers. While these documents are rich in detail, they can be difficult to navigate efficiently. To simplify the process of extracting relevant information from these extensive texts, {\modelname} combines RAG with LLMs, significantly enhancing the accuracy of document-related queries. The RAG framework consists of three key components: document indexing, retrieval, and generation. Initially, it processes external texts by breaking them down into manageable segments for the LLM's contextual analysis. It then creates and stores document embeddings for future retrieval. When a query is received, the system retrieves relevant embeddings to form a context window for the LLM, which then generates responses based on a prompt that includes both the query and the retrieved data.

{\modelname} adopts LangChain, an advanced open-source framework specifically designed for creating applications with LLMs, to implement RAG. LangChain's DocumentLoaders and Text Splitters are utilized to effectively organize and segment documents for query processing. Subsequently, VectorStore and Embeddings models are employed to generate and maintain document embeddings. For this task, we use \textit{all-MiniLM-L6-v2} document embeddings to create the embeddings and a FAISS-based similarity index vector storage for efficient retrieval. The Retriever component is crucial in obtaining the relevant segments to be included in the user-defined prompt. Lastly, the refined query, augmented with the retrieved data, is fed into {\llmname}, which generates customized responses. This demonstrates the smooth integration of RAG within {\modelname}, enhancing document comprehension in the field of scientific software.

\section{Case Study}

We deploy {\modelname} on the Energy Exascale Earth System Model (E3SM) as a case study to demonstrate its effectiveness in analyzing source code, code metadata, and text-based technical reports. 

\subsection{E3SM Introduction}
The E3SM is a state-of-the-art Earth system modeling framework, developed and supported by the United States Department of Energy (DOE)\cite{e3sm}. This model is designed to facilitate the detailed simulation of intricate interactions among the Earth's atmospheric, terrestrial, oceanic, and cryospheric components. Standing as one of the most advanced Earth system models globally, E3SM is extensively utilized by the scientific community to explore a large range of critical subjects, including climate dynamics, meteorological prediction, and the assessment of natural hazards. E3SM has a very long history and is written mainly by legacy programming languages such as Fortran. In addition, E3SM contains more than one million lines of code. Therefore, we think E3SM is a qualified large-scale scientific computing software for evaluating the proposed {\modelname}. Due to our special interest in kilometer-scale land component development on Exascale computers, we designed several demonstration queries focusing on the ultrahigh resolution E3SM Land Model (uELM)~\cite{uELM}.

\subsection{Source Code Query Results}
To illustrate the effectiveness of {\modelname} in source code analysis, we assessed its ability to generate FQL (Feature Query Language) queries across three distinct categories: Library Utilization Queries, Version Assessment Queries, and Feature Enumeration Queries. Through the use of crafted prompts, as demonstrated in the subsequent text boxes, {\modelname} successfully generates the correct FQL query in certain instances. Specifically, for the Library Utilization Query, {\modelname} produces the precise FQL query in a zero-shot manner. This accuracy is likely attributed to the pretraining of the LLaMA-2 model within {\modelname} on RAG supporting documents that include FQL queries and HPC terminologies. Similarly, for Version Assessment and Feature Enumeration Query, ``\textit{Please Generate FQL query to Find the minimum version of MPI}'' and ``\textit{Please Generate FQL query to List MPI process topology used in the Code}'' can prompt {\modelname} to generate accurate queries.

\begin{tcolorbox}
\small
    \textbf{Prompt for Library Utilization Query:} \\
    \textit{Please Generate FQL query to Check Whether OpenMP is used}\\[2mm] % Adjust vertical space after the prompt
    \textbf{Response:} \\
    \noindent
    \vspace{-0.7cm}
    \begin{verbatim}
CHECK (omp task || end task ||  omp taskloop || omp taskloop simd 
|| omp taskyield) WHERE (*)
    \end{verbatim}
    \vspace{-0.7cm}
\end{tcolorbox}

\begin{tcolorbox}
\small
    \textbf{Prompt for Version Assessment Query:} \\
    \textit{Please Generate FQL query to Find the minimum version of MPI}\\[2mm] % Adjust vertical space after the prompt
    \textbf{Response:} \\
    \noindent
    \vspace{-0.7cm}
    \begin{verbatim}
MAX (CHECK (MPI_AINT_DIFF) WHERE (*) AS (31), 
CHECK (MPI_COMM_DUP_WITH_INFO) WHERE (*) AS (30), 
CHECK (MPI_COMM_SET_INFO) WHERE (*) AS (30))
    \end{verbatim}
    \vspace{-0.7cm}
\end{tcolorbox}

\begin{tcolorbox}
\small
    \textbf{Prompt for Feature Enumeration Query:} \\
    \textit{Please Generate FQL query to List MPI process topology used in the Code}\\[2mm]
    \textbf{Response:} \\
    \noindent
    \vspace{-0.7cm}
    \begin{verbatim}
LIST (CHECK (MPI_CART_CREATE WHERE(*) AS (Cartesian), CHECK (MPI_GRAPH_CREATE WHERE(*) 
AS (Graph),  CHECK (MPI_DIST_GRAPH_CREATE_Adjacent WHERE(*) AS (Distributed Graph)), 
CHECK (omp parallel num_threads(dynamic) WHERE(*) AS (Dynamic Threads), 
CHECK (omp schedule(static, dynamic, guided, auto, runtime)) WHERE(*) AS (Scheduling))
    \end{verbatim}
    \vspace{-0.7cm}
\end{tcolorbox}

\subsection{Metadata Comprehension with {\modelname}}

In our study, we utilize three distinct forms of metadata to demonstrate the capabilities of {\modelname} in enhancing metadata comprehension. These include DOT files produced by XScan~\cite{zheng2019xscan}, CSV files generated by SPEL~\cite{schwartz2022spel}\footnote{A software toolkit designed for porting E3SM land model onto GPUs using OpenACC.}, and bespoke metadata formats also created by SPEL, all derived from the E3SM. This diverse dataset showcases {\modelname}'s versatility and effectiveness in interpreting and leveraging different metadata representations to deepen our understanding of complex scientific software systems.

Our analysis begins with assessing {\modelname}'s efficacy on standardized DOT files, as produced by XScan, and on a unique metadata format devised by SPEL. A DOT file, as illustrated in blow colored box, represents a directed graph that visualizes the interactions between caller and callee functions within E3SM, offering insights into the software's structural dynamics.

Structured query language (SQL) is widely used for processing information in relational databases, and {\llmname} models can conveniently translate natural language into SQL queries. To illustrate the SQL processing capabilities of {\modelname}, we employ two example CSV files generated by SPEL~\cite{schwartz2022spel}, as detailed in Table~\ref{tab:sql_tables}. 
Our approach involves a two-step process with {\modelname}. First, we instruct {\modelname} to generate SQL statements that transform the two subtables depicted in Table~\ref{tab:sql_tables} into formal SQL tables. 
Following this, we guide {\modelname} to generate specific SQL queries: first, to identify the name of the component characterized by a \textit{2D} \texttt{Dimension} and having a \texttt{Derived Type} of \textit{col\_pp}; and second, to construct a new SQL view by joining these two subtables. This methodology underscores {\modelname}'s adeptness at navigating and manipulating SQL tables, showcasing its proficiency in facilitating advanced database operations.

\begin{table}[h]
\centering
\caption{Two CSV files produced by SPEL~\cite{schwartz2022spel} to be processed as SQL tables}
\label{tab:sql_tables}

\begin{subtable}{0.49\textwidth}
\centering
\caption{Detailed information about \texttt{Component}}
\label{tab:subtable1}
    \begin{tabular}{|l|l|l|}
    \hline
    \textbf{Variable} & \textbf{Type} & \textbf{Dimension} \\ \hline
    snl                & integer       & 1D                 \\ \hline
    dz                 & real          & 2D                 \\ \hline
    sabg\_patch        & real          & 1D                 \\ \hline
    sabg\_lyr\_patch   & real          & 2D                 \\ \hline
    ws\_col            & real          & 1D                 \\ \hline
    lake\_icefrac\_col & real          & 2D                 \\ \hline
    \end{tabular}
\end{subtable}
% \vspace{1em} % Add some space between the subtables
\begin{subtable}{0.49\textwidth}
\centering
\caption{\texttt{Derived Types} of each \texttt{Component}}
\label{tab:subtable2}
    \begin{tabular}{|l|l|}
    \hline
    \textbf{Derived Type} & \textbf{Component} \\ \hline
    col\_pp               & snl                \\ \hline
    col\_pp               & dz                 \\ \hline
    solarabs\_vars        & sabg\_patch        \\ \hline
    solarabs\_vars        & sabg\_lyr\_patch   \\ \hline
    lakestate\_vars       & ws\_col            \\ \hline
    lakestate\_vars       & lake\_icefrac\_col \\ \hline
    \end{tabular}
\end{subtable}

\end{table}

\begin{tcolorbox}
\small
    \textbf{Prompt for Summarizing DOT File:} \\
    \textit{DOT file can describe a directed graph using keyword of  ``digraph". Inside a directed graph, each line starts with a module name which is followed by the ``::" symbol. For example, ``allocationmod" is the module in ``allocationmod::allocationinit -$>$ clm\_time\_manager." Please list all the unique modules in the following DOT file describing directed graph G.}
    \begin{verbatim}
digraph G {
    emi_datamod::emidcopy -> emi_datamod::emidallocatememory_real_3d
    emi_datamod::emidcopy -> emi_datamod::emidallocatememory_real_4d
    elm_driver::elm_drv -> decompmod::get_proc_clumps
    elm_driver::elm_drv -> satellitephenologymod::interpmonthlyveg
    elm_driver::elm_drv -> activelayermod::alt_calc
    elm_driver::elm_drv -> verticalprofilemod::decomp_vertprofiles
    elm_driver::elm_drv -> balancecheckmod::begingridwaterbalance
    elm_driver::elm_drv -> dynsubgriddrivermod::dynsubgrid_driver
    elm_driver::elm_drv -> ndepstreammod::ndep_interp
    ch4varcon::ch4conrd -> fileutils::relavu
    elm_driver::elm_drv -> firemod::fireinterp
    canopyhydrologymod::canopyhydrology_readnl -> fileutils::relavu
    elm_driver::elm_drv -> pdepstreammod::pdep_interp
    controlmod::control_init -> fileutils::relavu
    firemod::firefluxes -> elm_nlutilsmod::find_nlgroup_name
    firemod::firefluxes -> ndepstreammod::elm_domain_mct
    firemod::firefluxes -> histfilemod::hist_addfld1d
}
    \end{verbatim}
    \vspace{-0.4cm}
    \textbf{Response:} \\
    \noindent
    % \vspace{-0.7cm}
Sure! Here are the unique modules in the directed graph G described by the DOT file:
\\
\texttt{1. emi\_datamod \\
2. elm\_driver \\
...\\
16. histfilemod\\
17. verticalprofilemod 
} \\
There are 17 unique modules in total.
    % \vspace{-0.7cm}
\end{tcolorbox}

\begin{tcolorbox}
\small
    \textbf{Prompt for Creating SQL Views:} \\
    \textit{Please generate SQL commands to construct a new SQL view by joining the following two SQL tables:: \texttt{Table 1 Info}; \texttt{Table 2 Info}. }
    \vspace{0.1cm}
    \\
    \textbf{Response:}
    \noindent
    \vspace{-0.3cm}
    \begin{verbatim}
CREATE VIEW myview AS
SELECT t1.Component, t1.Type, t1.Dimension, t2.DerivedType
FROM table1 t1
JOIN table2 t2 ON t1.Component = t2.Component;
SELECT * FROM myview;
    \end{verbatim}
    \vspace{-0.7cm}
\end{tcolorbox}

\begin{tcolorbox}
\small
    \textbf{Prompt for Creating SQL Table from CSV:} \\
    \textit{Please generate SQL commands to covert the following textural table into an SQL table with \texttt{Component} as the PRIMARY KEY: }
    \vspace{-0.2cm}
    \begin{verbatim}
Component,Type,Dimension
snl,integer,1D
dz,real,2D
sabg_patch,real,1D
sabg_lyr_patch,real,2D
ws_col,real,1D
lake_icefrac_col,real,2D
    \end{verbatim}
    \vspace{-0.5cm}
    \textbf{Response:}
    \noindent
    \vspace{-0.3cm}
    \begin{verbatim}
CREATE TABLE components (
Component VARCHAR(255) PRIMARY KEY,
Type VARCHAR(255),
Dimension VARCHAR(255)
);

INSERT INTO components (Component, Type, Dimension) VALUES
(`snl', `integer', `1D'),
(`dz', `real', `2D'),
(`sabg_patch', `real', `1D'),
(`sabg_lyr_patch', `real', `2D'),
(`ws_col', `real', `1D'),
(`lake_icefrac_col', `real', `2D');
    \end{verbatim}
    \vspace{-0.7cm}
\end{tcolorbox}

\begin{tcolorbox}
\small
    \textbf{Prompt for Querying SQL Table:} \\
    \textit{Please generate SQL commands to identify the name of the component characterized by a \textit{2D} \texttt{Dimension} and having a \texttt{Derived Type} of \textit{col\_pp} by joining the following two SQL tables: \texttt{Table 1 Info}; \texttt{Table 2 Info}. }
    \vspace{0.1cm}
    \\
    \textbf{Response:}
    \noindent
    \vspace{-0.3cm}
    \begin{verbatim}
SELECT Component
FROM Table1
JOIN Table2 ON Table1.Component = Table2.Component
WHERE Dimension = `2D' AND DerivedType = `col_pp';
    \end{verbatim}
    \vspace{-0.7cm}
\end{tcolorbox}

Additionally, a sequence of execution for \texttt{LakeTemperature} as delineated by SPEL~\cite{schwartz2022spel} is shown in the colored box below. This example highlights the execution sequence within E3SM, where the first column lists variables utilized by the \texttt{LakeTemperature} module. The top row segments the execution flow into distinct phases \{main (\texttt{LakeTemperature}) -$>$ subroutine -$>$ main -$>$ subroutine -$>$ main\}, separated by ``$|$", depicting the procedural pathway. Each column within these sections signifies a Fortran ``do loop" sequence. The entries within this table specify each variable's role in the corresponding ``do loop", with potential values being ``ro" (read-only), ``wo" (write-only), ``rw" (read-write), or ``-" (not in use). This offers an in-depth view of variable usage during the execution process, which is beneficial for code performance optimization (through asynchronous kernel launch) on GPUs. This structured approach to metadata analysis underscores {\modelname}'s capability to navigate and elucidate complex metadata representations, enhancing comprehension and facilitating deeper insights into the software's operational mechanisms.

\begin{tcolorbox}
\small
    \textbf{Prompt for Analyzing Metadata of \texttt{LakeTemperatureAllLoopVariables} generated by SPEL~\cite{schwartz2022spel}:} \\
    \textit{Here is the data format information for the LakeTemperatureAllLoopVariables.txt file: 
1. The first column contains the name of variables that are used by the LakeTemperature.
2. The top row displays various sections (divided by `` $|$") that illustrate the sequence of execution:
lakeTemperature(main) -$>$ its subroutine->return to main -$>$ its subroutine -$>$ main
3. Every column within code section (divided by `` $|$") represents a ``do loop" (Fortran) segment.
The value of each table element indicates the usage of a variable in that ``do loop", the value could be ``ro, wo, ro, -". ro: read-only, wo: write-only, rw: read-write, -: not in use"\\
The content of LakeTemperatureAllLoopVariables.txt file:
}
\footnotesize{
    \begin{verbatim}
                   |LakeTemperature        |soilthermprop_|LakeTemperature          |
filter_lakec       |ro ro -  ro ro ro -  ro|ro ro ro ro   |ro ro -  ro ro - ro ro ro|
lake_col_to_filter |wo -  ro -  -  -  ro - |- -  -  -     |-  -  ro -  -  -  -  -  -|
ocvts              |wo -  -  -  -  -  -  - |- -  -  -     |rw rw -  -  -  -  -  -  -|
puddle             |wo -  -  -  -  -  -  - |- -  -  -     |-  -  -  -  -  -  -  -  -|
frzn               |wo -  -  -  -  -  -  - |- -  -  -     |-  -  -  -  -  -  -  -  -|
bottomconvect      |wo -  -  -  -  -  -  - |- -  -  -     |-  -  -  -  -  -  -  -  -|
hc_soisno          |wo -  -  -  -  -  -  - |- -  -  -     |-  -  -  -  -  -  -  -  -|
hc_soi             |wo -  -  -  -  -  -  - |- -  -  -     |-  -  -  -  -  -  -  -  -|
    \end{verbatim}}
    \textit{\textbf{How many variables are used by the first do loop in lakeTemparature. And what is the role of these variables (ro, wo, or rw)?}} \\
    \textbf{Response:} \\
    \noindent
This indicates that there are 8 variables used in the first ``do loop" of the LakeTemperature section. Their roles are as follows:\\
- 1 variable is read-only (filter\_lakec).\\
- 7 variables are write-only (lake\_col\_to\_filter, ocvts, puddle, frzn, bottomconvect, hc\_soisno, hc\_soi).
    % \vspace{-0.7cm}
    }
\end{tcolorbox}

\subsection{Document Parsing}

This section demonstrates the effectiveness of combining the RAG framework and LangChain with LLMs, as described in Section~\ref{sec:doc-rag}, for extracting specific technical information from comprehensive scientific documents. We utilize a 28-page technical report on Lake Temperature (in PDF format), from the Community Land Model~\cite{clm4.5}, as the testbed. In Experiment 1, \modelname adeptly retrieves information regarding the calculation of thermal conductivities at layer interfaces, showcasing the method's ability to understand and extract specific computational techniques, specifically the harmonic mean approach. Experiment 2 further demonstrates \modelname's efficacy by accurately describing the application of the Crank-Nicolson method within the model, showcasing its competence in elucidating detailed methodological implementations. Lastly, Experiment 3 confirms \modelname's proficiency in extracting complex physical processes, such as phase changes, and its capability in both comprehension and explanation. Collectively, these experiments underscore the robustness of our approach in navigating and interpreting sophisticated scientific documents, affirming its potential to significantly enhance the comprehension of complex scientific documentation interpretation.

\begin{tcolorbox}
\small
    \textbf{Prompt for Querying Technical Description of Lake Temperature -1:} \\
    \textit{What method is used to calculate the thermal conductivities at the interfaces between layers in the model?}
    \vspace{0.1cm}
    \\
    \textbf{Response:}
    \noindent
    % \vspace{-0.3cm}
The thermal conductivities at layer interfaces are calculated as the harmonic mean of the conductivities of the neighboring layers.
    % \vspace{-0.7cm}
\end{tcolorbox}

\begin{tcolorbox}
\small
    \textbf{Prompt for Querying Technical Description of Lake Temperature -2:} \\
    \textit{How is the Crank-Nicolson method applied in the context of this model?}
    \vspace{0.1cm}
    \\
    \textbf{Response:}
    \noindent
    % \vspace{-0.3cm}
    The Crank-Nicolson method is used to solve equation (9.44) and it results in a tridiagonal system of equations.
    % \vspace{-0.7cm}
\end{tcolorbox}

\begin{tcolorbox}
\small
    \textbf{Prompt for Querying Technical Report of Lake Temperature -3:} \\
    \textit{What is the significance of phase changes in the model, and how are they accounted for?}
    \vspace{0.1cm}
    \\
    \textbf{Response:}
    \noindent
    Phase changes, such as freezing and melting, play a significant role in the atmosphere model, as they affect the energy balance at the land surface. The model accounts for phase changes by calculating heat diffusion and phase change in given layers based on temperature and moisture conditions. The available energy for melting is computed as function of temperature and moisture content.
\end{tcolorbox}

\section{Discussion and Future Work}

The foundational components of \textit{\modelname}, LLMs, have demonstrated remarkable proficiency in interpreting various data sources, including source code, metadata, and documents, offering insights from multiple angles. Despite these achievements, there are inherent limitations within LLMs that warrant further investigation in our future work. First, the constraint related to the maximum token count supported by open-source LLMs necessitates expansion to accommodate extensive queries, aiming to enhance in-context learning capabilities. Moreover, the treatment of specialized terminologies within domain-specific sciences requires refinement to ensure more accurate and dependable processing. Currently, commercial versions like GPT-4 are employed in {\modelname} to derive coding keywords from terminologies; however, we posit that leveraging a more transparent, reproducible model, amenable to fine-tuning for domain-specific sciences, could optimize this process. Furthermore, {\modelname} could be enriched with additional functionalities, such as identifying and addressing computational bottlenecks within large-scale codes and providing direct source code optimization recommendations. These areas of potential enhancement motivate us for the continued development and improvement of {\modelname} in future iterations.

\section{Conclusion}
This paper presents {\modelname}, a framework developed on the foundation of LLMs, aimed at unraveling the complexities inherent in large-scale scientific software. By enhancing the capabilities of {\llmname} models within {\modelname} through innovative approaches such as instruction-based prompting, integration of external GPT-4 queries, and the adoption of Retrieval-Augmented Generation (RAG) and LangChain techniques, we have significantly expanded the operational capacity of pre-trained LLMs. Our comprehensive evaluation across a variety of data types, including source code, diverse metadata formats (DOT, SQL, and specialized schemas), and textual documents, has validated the efficacy of {\modelname}. It is our aspiration that {\modelname} will illuminate pathways for forthcoming inquiries in the fields of generative AI and software engineering, particularly within the domain of scientific computing.

\bibliographystyle{plain}
\bibliography{reference}

\end{document}